\documentclass[aps,twocolumn,amssymb,amsmath,superscriptaddress]{revtex4-1}

\usepackage{bm}
\usepackage{color}
\usepackage{graphicx}
\usepackage{dcolumn}
\usepackage{graphicx}
\usepackage[colorlinks=true, letterpaper=true, pdfstartview=FitV, linkcolor=blue, citecolor=blue, urlcolor=blue]{hyperref}
\usepackage[normalem]{ulem}
\usepackage{amsmath}
\usepackage{epstopdf}
\usepackage{multirow}

\setcitestyle{square}

\makeatletter
\renewcommand\@biblabel[1]{#1.}
\makeatother

\begin{document}

\title{Ferroelectric Antiferromagnetic Lifting of Spin-Valley Degeneracy}

\author{Jiaqi Feng}
\affiliation{Laboratory of Quantum Functional Materials Design and Application, School of Physics and Electronic Engineering, Jiangsu Normal University, Xuzhou 221116, China}

\author{Xiaodong Zhou}
\email{zhouxiaodong@tiangong.edu.cn}
\affiliation{School of Physical Science and Technology, Tiangong University, Tianjin 300387, China}
\affiliation{Laboratory of Quantum Functional Materials Design and Application, School of Physics and Electronic Engineering, Jiangsu Normal University, Xuzhou 221116, China}

\author{Jingyan Chen}
\affiliation{Laboratory of Quantum Functional Materials Design and Application, School of Physics and Electronic Engineering, Jiangsu Normal University, Xuzhou 221116, China}

\author{Meiling Xu}
\affiliation{Laboratory of Quantum Functional Materials Design and Application, School of Physics and Electronic Engineering, Jiangsu Normal University, Xuzhou 221116, China}

\author{Xiuxian Yang}
\affiliation{Laboratory of Quantum Functional Materials Design and Application, School of Physics and Electronic Engineering, Jiangsu Normal University, Xuzhou 221116, China}

\author{Yinwei Li}
\email{yinwei$_$li@jsnu.edu.cn}
\affiliation{Laboratory of Quantum Functional Materials Design and Application, School of Physics and Electronic Engineering, Jiangsu Normal University, Xuzhou 221116, China}

\date{\today}

\begin{abstract}
The generation and control of spin- and valley-polarization in antiferromagnets (AFMs) have garnered increasing attention due to their potential for enabling faster and more stable multifunctional spintronic and valleytronic memory and logic devices. However, the two primary categories of AFMs $-$ altermagnets and $\mathcal{TP}$-symmetric AFMs $-$ either lack intrinsic valley-polarization or net spin-polarization. Here, we propose an effective approach for achieving spontaneous spin-valley polarization in $\mathcal{TP}$-broken layered ferroelectric antiferromagnets (FE-AFMs). The FE-AFMs exhibit lifted spin degeneracy across the entire Brillouin zone, along with uncompensated spin density of states. They combine the benefits of spin-polarization in altermagnets with valley-polarization in $\mathcal{TP}$-symmetric AFMs.  Furthermore, the FE-AFMs feature layer-dependent spin-polarization, rooted in their intrinsic ferroelectric property, allowing for the flexible control over spin-valley polarization by interlayer sliding. This tunability facilitates sign-reversible and size-tunable valley Hall and Nernst effects, along with other spin-valley-dependent transport properties. Our findings are demonstrated in a broad class of $\mathcal{TP}$-broken bilayer antiferromagnets such as Nb$_3X_8$ ($X$ = Cl, Br, I), V$X_2$ ($X$ = S, Se), and VSi$_2X_4$ ($X$ = N, P), underscoring the potential of FE-AFMs for advancing next-generation spin- and valley-based information technologies.
\end{abstract}

\maketitle

\section{Introduction}\label{intro}

The central theme of spintronics is the active manipulation of spin degree of freedom,  which serves as an ideal carrier for non-volatile information storage and processing in solid-state systems~\cite{Zutic2004,Chappert2007}. Spin polarization of electrons with asymmetric spin states set the foundation for various advanced spintronic applications, including anomalous and spin transports~\cite{YG-Yao2004,Nagaosa2010,Sinova2015,XD-Zhou2022}, magneto-optical effects~\cite{Ebert1996,WX-Feng2015,XD-Zhou2019a,XD-Zhou2021}, spin-torque phenomena~\cite{Zelezny2018,Hernandez2021,Bose2022,H-Bai2022,H-Bai2023,Karube2022,DF-Shao2023}, giant and tunneling magnetoresistance~\cite{DF-Shao2021,Smejkal2022c}. Meanwhile, valley degrees of freedom, referring to the multiple degenerate energy extrema at the conduction or valence band edges, are emerging as promising information carriers for valleytronics~\cite{XD-Xu2014,Schaibley2016,Mak2018}. A primary focus in valleytronics is to lift valley degeneracy, thereby enabling valley-related transport phenomena such as valley Hall effect (VHE)~\cite{D-Xiao2007,D-Xiao2012,Lee2016} and valley Nernst effect (VNE)~\cite{XQ-Yu2015,Dau2019}. For multidisciplinary applications, achieving combined spin-valley polarization is particularly desirable.

\begin{figure}
	\includegraphics[width=1\columnwidth]{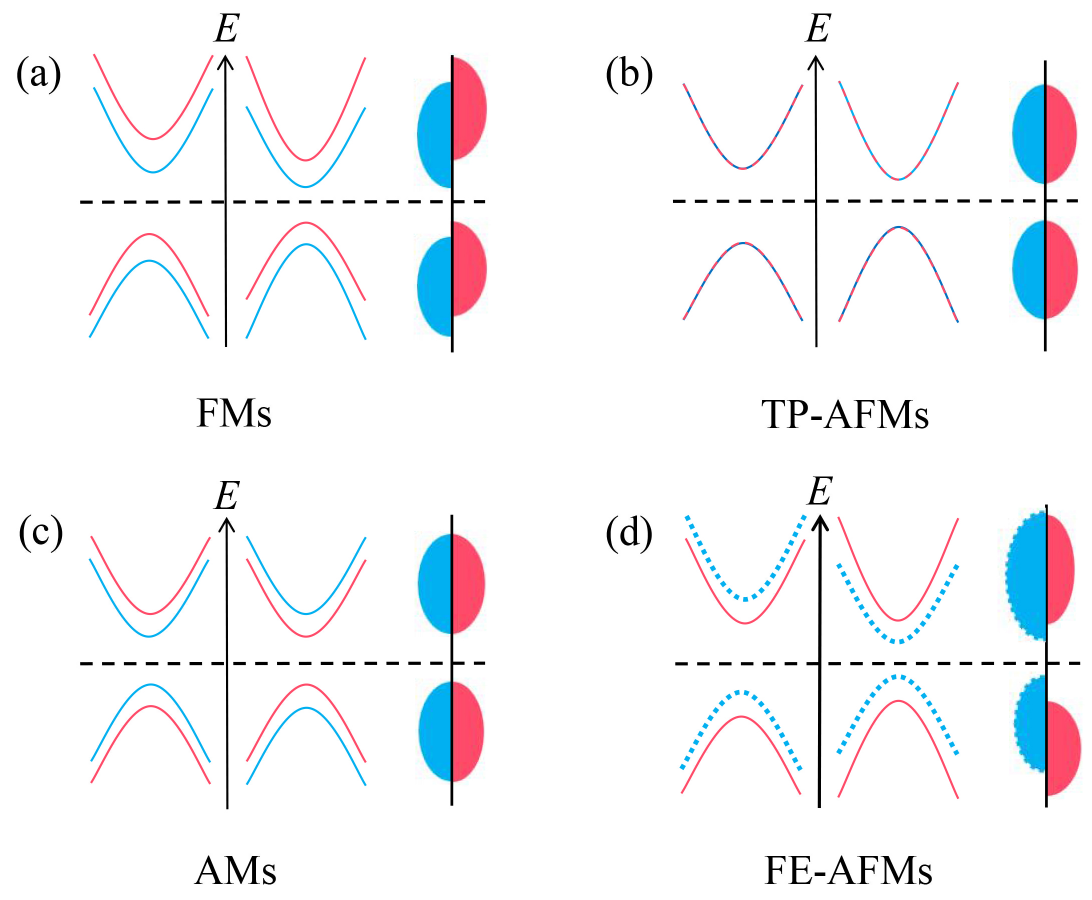} 
	\caption{(Color online) Schematic illustration of spin-valley characteristics in four types of magnetic materials. (a) Spin-valley-splitting and uncompensated DOS in FMs. (b) Spin degeneracy, valley-splitting, and compensated DOS in conventional $\mathcal{TP}$-AFMs. The solid and dashed lines represent spin-up and spin-down states, respectively. (c) Alternating and local spin-splitting, valley degeneracy, and compensated DOS in AMs. (d) Global spin-splitting, valley-splitting, and uncompensated DOS in $\mathcal{TP}$-broken FE-AFMs. Solid red and dashed blue lines represent spin-up and spin-down states for the bottom and top layers, respectively. The corresponding solid blue (spin-down) and dashed red lines (spin-up) from the bottom and top layers, are located away from the Fermi level and are not shown.}
	\label{fig:map}
\end{figure}

Ferrovalley materials~\cite{WY-Tong2016} represent the first intrinsic candidates capable of exhibiting spontaneous spin-valley polarization without external magnetic or optical fields. This spin-valley polarization arises from the interplay between ferromagnetism and spin-orbit coupling (SOC), as shown in Fig.~\ref{fig:map}(a). Recent research interest in spin-valley polarization has been steadily extending from ferromagnets (FMs) to antiferromagnets (AFMs) due to its minimized stray fields and ultrafast spin dynamics~\cite{Smejkal2018,Jungwirth2018,Baltz2018}. Two types of AFMs have been proposed for hosting spin-valley physics, that are, the $\mathcal{TP}$-symmetric AFMs ($\mathcal{TP}$-AFMs) and $\mathcal{TP}$-broken altermagnets (AMs)~\cite{Smejkal2022a,Smejkal2022b,XD-Zhou2024,L-Bai2024,YC-Liu2024} ($\mathcal{T}$ is time-reversal; $\mathcal{P}$ is spatial inversion). The former exhibits spontaneous valley-polarization but lacks spin-polarization~\cite{X-Li2013}, as shown in Fig.~\ref{fig:map}(b). The spin-degeneracy can be lifted under a vertical electric field by breaking the $\mathcal{TP}$ symmetry~\cite{Sivadas2016}. In comparison, the altermagnets feature alternating spin-polarization at different valleys which are related to each other by a crystal symmetry rather than conventional $\mathcal{T}$ symmetry~\cite{HY-Ma2021,RW-Zhang2024}, as shown in Fig.~\ref{fig:map}(c). The valley-polarization is intrinsically absent but it can be excited by a gate-electric field or strain field~\cite{HY-Ma2021,RW-Zhang2024}. Both types of AFMs harbor a compensated density of state without net spin-polarization. Exploring AFMs that exhibit both spontaneous spin- and valley-polarization holds promise for advancing antiferromagnetic spintronic and valleytronic applications.

Here, we propose a novel platform for achieving spontaneous antiferromagnetic spin-valley polarization in $\mathcal{TP}$-broken layered ferroelectric antiferromagnets (FE-AFMs). This approach combines the benefits of valley-polarization in $\mathcal{TP}$-AFMs and spin-polarization in AMs. Unlike $\mathcal{TP}$-AFMs and AMs, which exhibit either zero or localized spin-polarization with compensated density of states (DOS), the FE-AFMs are characterized by lifted spin degeneracy across the entire Brillouin zone and an uncompensated DOS  [See Fig.~\ref{fig:map}(d)].  Another distinguishing feature of FE-AFMs is their layer-dependent spin-polarization, where distinct spin components polarized in different layers, a property rooted in the intrinsic out-of-plane ferroelectric polarization. Moreover, the FE-AFMs offer the unique advantage of enabling flexibly control both the signs and magnitudes of spin-valley polarization via interlayer sliding. This tunability inevitably enables the sign-reversible and size-tunable VHE and VNE. We further uncover the switchable spin-valley polarization as well as the valley-dependent transport properties are present in a broad class of $\mathcal{TP}$-broken layered FE-AFMs, such as Nb$_3X_8$ ($X=$ Cl, Br, I), V$X_2$ ($X$ = S, Se), VSi$_2X_4$ ($X=$ N, P). Our findings position FE-AFMs as an ideal platform for spin- and valley-based antiferromagnetic devices, leveraging the unique capabilities of slidetronics and ferroelectrics~\cite{L-Li2017,Sivadas2018}.

\section{Results and discussion}\label{results}

\begin{figure}[ht]
	\includegraphics[width=1\columnwidth]{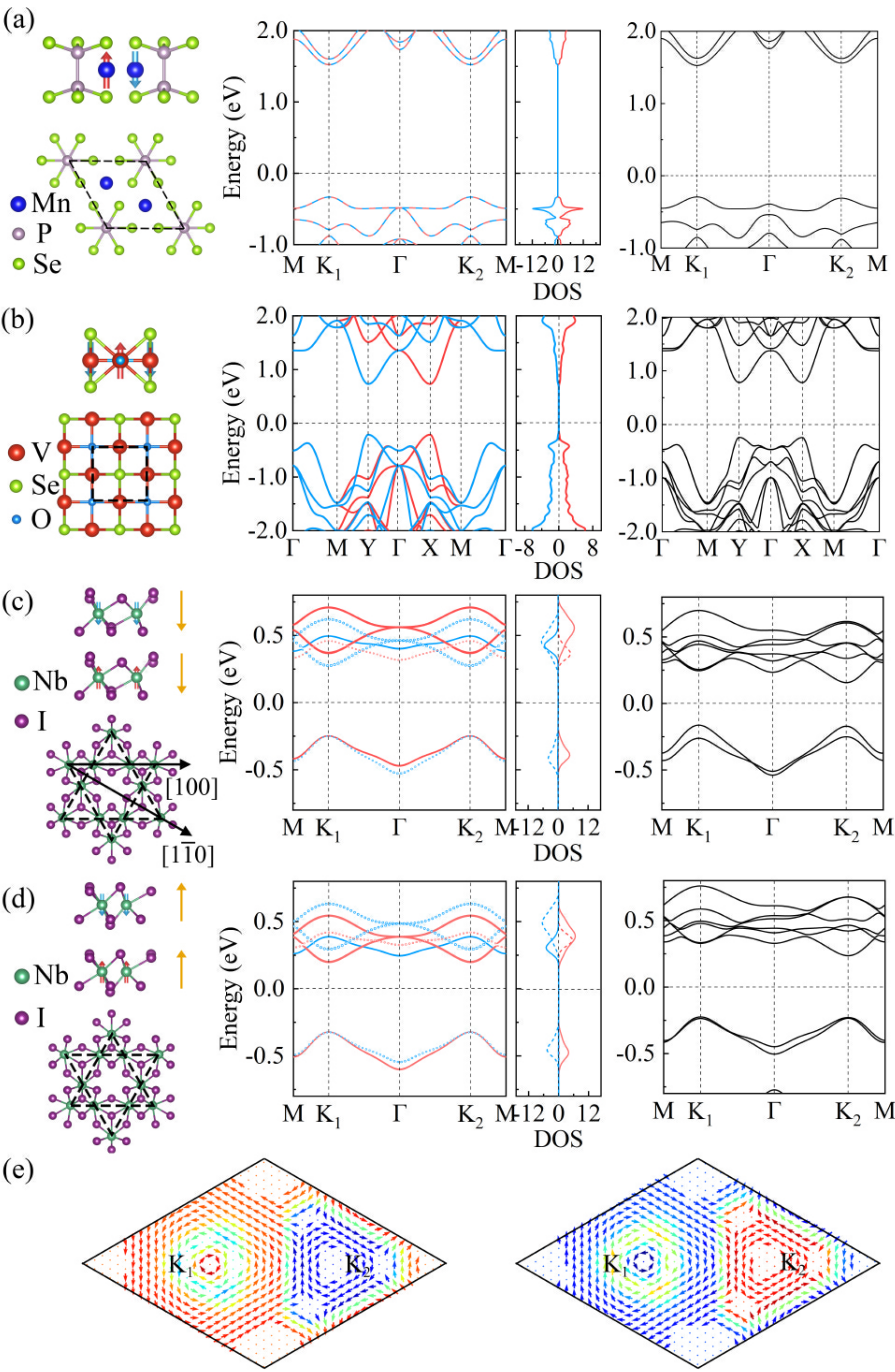}
	\caption{(Color online) Spin-valley characteristics in three types of specific AFMs. (a)-(b) Side and top views of crystal structure, spin-polarized band structure and DOS, and relativistic band structures in monolayer $\mathcal{TP}$-AFM MnPSe$_3$ and AM V$_2$Se$_2$O, respectively. (c)-(d) Side and top views of crystal structure, layer-resolved spin-polarized band structures and DOS, and relativistic band structures in bilayer antiferromagnetic Nb$_3$I$_8$ with opposite ferroelectric polarization (i.e., AA and BB stacking), respectively. The solid and dashed lines indicate spin contributions from the bottom and top layers, respectively. The black arrows in (c) indicate the sliding direction along [100] and [1$\bar{1}0$] directions. The red and blue arrows in (a)-(d) indicate two opposite spin sublattices. The orange arrows in (c)-(d) represent the direction of ferroelectric polarization. (e) Spin textures at the bottom of the conduction band in Nb$_3$I$_8$ with opposite ferroelectric polarization, where in-plane spin components shown by arrows and out-of-plane components represented by colors. }
	\label{fig:structure}
\end{figure}

Before discussing spin-valley polarization in FE-AFMs, we first briefly review the existing paradigms in $\mathcal{TP}$-AFMs and AMs, using monolayer MnPSe$_3$ and V$_2$Se$_2$O as representative examples. Fig.~\ref{fig:structure}(a) illustrates the magnetic structure and electronic properties of MnPSe$_3$, which exhibits a 
N{\'e}el-type antiferromagnetic order with a magnetic point group of $\bar{3}^{\prime}$. The $\mathcal{TP}$ operation in this group enforces Kramers spin degeneracy, resulting in spin-compensated DOS. But the valley degeneracy is lifted due to the broken $\mathcal{T}$ and $\mathcal{P}$ symmetries~\cite{X-Li2013}. Consequently, spin-polarization is forbidden, while valley-polarization is allowed in $\mathcal{TP}$-AFMs. In contrast, AMs exhibit the opposite behavior. Fig.~\ref{fig:structure}(b) depicts the magnetic structure and electronic properties of V$_2$Se$_2$O, where the two sublattices are related by a spin group operation [$\mathcal{C}_2||M_{\bar{1}10}$]. Here, $\mathcal{C}_2$ acts solely in spin space that changes the spin sign, while $M_{\bar{1}10}$ is a mirror operation that acts solely in real space. Because [$\mathcal{C}_2||M_{\bar{1}10}$]$E(k_x, k_y, k_z, s)$ = $E(k_y, k_x, k_z, -s)$, the spin group symmetry ensures that the two valleys are energetically degenerate but carry alternating spin orientations, leading to the novel $C$-paired spin-valley locking~\cite{HY-Ma2021}. As a result, local spin-polarization is permitted, while valley-polarization is prohibited in AMs without external means, accompanying a compensated DOS.

Next, we focus on FE-AFMs, which satisfy the desired criteria of antiferromagnetic spin-valley-polarization and uncompensated DOS. The FE-AFMs can be easily designed by stacking two ferrovalley monolayers. The typical candidates are Nb$_3X_8$ ($X=$ Cl, Br, I)~\cite{R-Peng2020,Conte2020,Cantele2022,L-Feng2023,YL-Feng2023,JY-Duan2024}, 2H-V$X_2$ ($X$ = S, Se)~\cite{WY-Tong2016,XG-Liu2020}, and VSi$_2X_4$ ($X=$ N, P)~\cite{XD-Zhou2021b,QR-Cui2021,S-Li2021,XY-Feng2021}, as demonstrated in Appendix~\ref{appendix2} Fig.~\ref{fig:FIG_S1}. In the following discussion, we take the bilayer Nb$_3$I$_8$ as an example to explore the spin-valley polarization in FE-AFMs, and discuss the rest of the materials in the Appendix~\ref{appendix2} (see Fig.~\ref{fig:FIG_S2}). Recent theoretical and experimental studies have highlighted the rich layer-dependent magnetic and ferroelectric properties in Nb$_3$I$_8$~\cite{JK-Jiang2017,R-Peng2020,Conte2020,Cantele2022,L-Feng2023,YL-Feng2023,J-Hong2023,JY-Duan2024}. Fig.~\ref{fig:structure}(c) depicts its magnetic structure and electronic properties. The AA-stacked Nb$_3$I$_8$ exhibits spontaneous ferroelectric polarization along the out-of-plane direction, arising from its unique breathing magnetic kagome lattice shaped by Nb trimers combined with  distorted I octahedron environment. This unique breathing-type ferroelectricity has also been demonstrated in other trimer-based compounds such as Ta$_3$I$_8$~\cite{SC-Xing2024,JJ-Lu2024} and W$_3$Cl$_8
$~\cite{D-Hu2024,YB-Liu2024}. The structure favors an A-type antiferromagnetic ground state with a magnetic point group of 3$m^{\prime}$1. Under this group, the two spin sublattices are unrelated by any symmetry operations, leading to the simultaneous lifting of spin and valley degeneracy. As a result, a net spin-resolved DOS can be observed even in the absence of net magnetization, which is a defining feature of FE-AFMs.

We note that the system studied here, hosts antiparallel and compensated magnetic moments on two magnetic sublattices that are not related by any symmetry operation. Such a state can also be classified as a compensated ferroelectric ferrimagnet based on the symmetry consideration~\cite{Mazin2022,LD-Yuan2024,YC-Liu2025,SD-Guo2025,PJ-Guo2025}. The total magnetization is expressed as
\begin{equation}
M = \mu_{B}(N_{\uparrow}-N_{\downarrow})
\end{equation}
where $N_{\uparrow}$ and $N_{\downarrow}$ denote the numbers of spin-up and spin-down electrons, respectively~\cite{YC-Liu2025}. In bilayer Nb$_3$I$_8$, these are equal, resulting in full compensation of spin moments. Despite the absence of net magnetization, the inequivalence of the magnetic sublattices breaks the combined $\mathcal{TP}$ symmetry, giving rise to spin-valley polarization. In line with terminology increasingly adopted in recent literature on two-dimensional magnetoelectric materials~\cite{SQ-Du2020,AY-Gao2021,R-Chen2024,FR-Yao2025,XG-Liu2020,SD-Guo2025b}, we refer to this phase as a FE-AFMs, emphasizing the coexistence of ferroelectric order and fully compensated antiferromagnetism.

Furthermore, because the two spin sublattices are located in different layers, the spin-polarized bands exhibit layer-dependent spin-splitting. For example, near the Fermi level, the first conduction band and the second valence band with spin-down states are contributed by the upper layer, while the second conduction band and the first valence band with spin-up states are contributed by the bottom layer. These layer-polarized electronic states arise from the out-of-plane ferroelectric polarization~\cite{XG-Liu2020,P-Jiang2020,AY-Gao2021,YY-Feng2023,TF-Cao2023}. Reversing the direction of the ferroelectric polarization (i.e., from AA to BB stacking by a $\mathcal{TP}$ operation) leads to corresponding reversal in two spin states, with contributions from opposite layer spin channels, as shown in Fig.~\ref{fig:structure}(d). The spin textures of the two polarization states, depicted in Fig.~\ref{fig:structure}(e), clearly demonstrate the asymmetric spin distributions at the two valleys, which reverse when the polarization direction is switched. Similar layer-dependent electronic states have also been observed in other layered systems~\cite{SD-Guo2023,SD-Guo2024,W-Xun2024}.

\begin{figure*}
	\includegraphics[width=2\columnwidth]{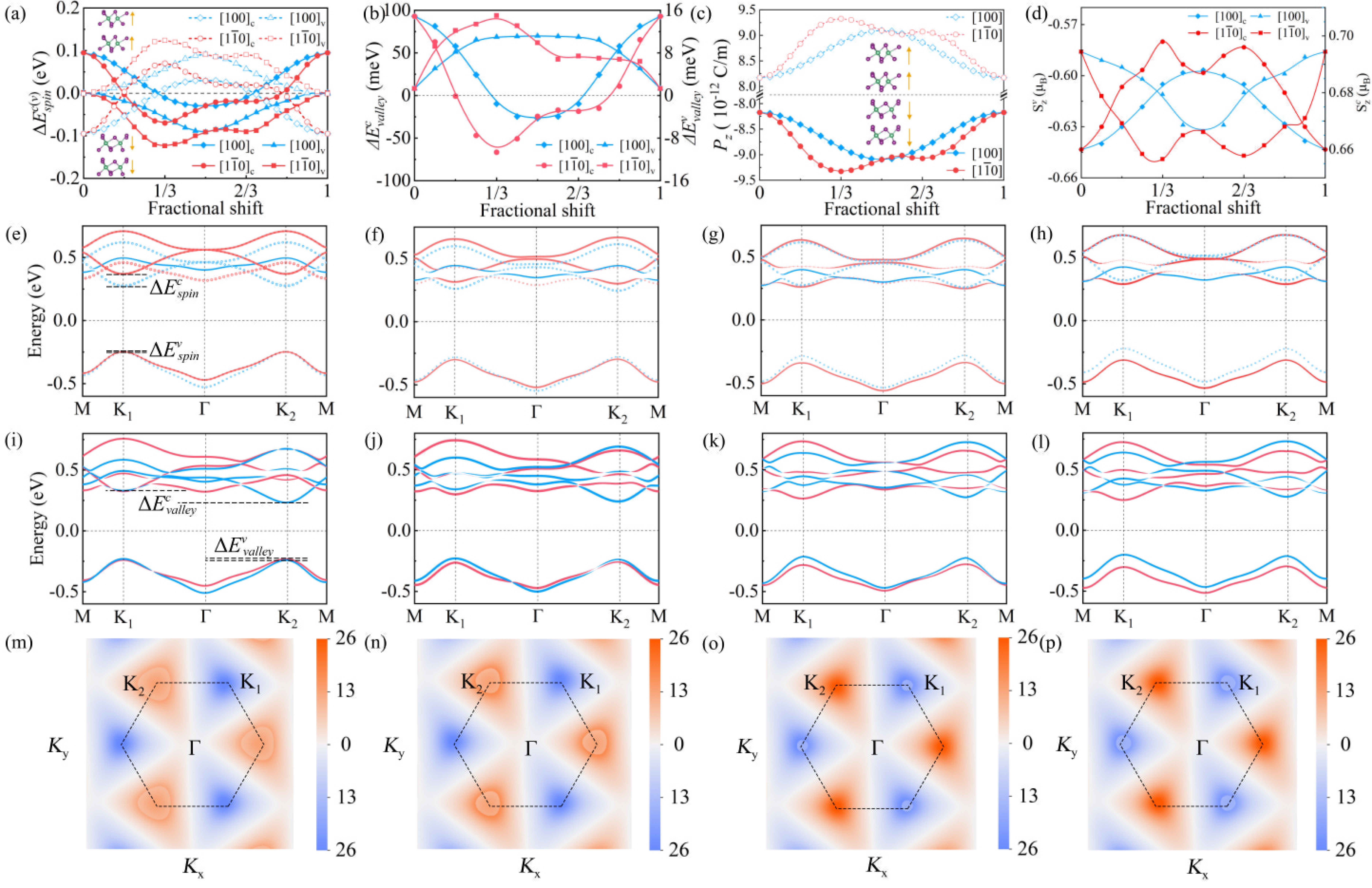}
	\caption{(Color online) Slidetronic Spin-Valley Polarization and Ferroelectric Polarization. (a-d) Spin-splitting, valley-splitting, out-of-plane ferroelectric polarization component, and spin polarization of the valence band maximum $s_{z}^{v}$  and conduction band minimum $s_{z}^{c}$ at the K$_1$ valley as a function of lateral shift along the [100] (blue line) and $[1\bar{1}0]$ (red line) directions.  In panels (a) and (b), the subscript $v$ and $c$ denote the spin- and valley-splitting at the valance and conduction band edges, respectively, as defined in Fig.~\ref{fig:splitting}(e) and~\ref{fig:splitting}(i). The solid and dot lines in panels (a) and (c) represent ferroelectric polarization along $-z$ and $z$ direction, respectively. Panels (e-h), (i-l), and (m-p) display the layer-resolved spin-polarized band structures, spin-resolved relativistic band structures, and Berry curvature with 0, $\frac{1}{6}$, $\frac{1}{3}$ and $\frac{1}{2}$ fractional shift along the [100] direction, respectively. The Berry curvature in Fig. 3(m-p) was evaluated at an energy between the conduction band minima of the two valleys, where nonzero transport contributions arise.}
	\label{fig:splitting}
\end{figure*} 

Interestingly, the layer-dependent spin-valley polarization can be effectively switched through interlayer sliding. To investigate this, we calculated the stacking energy and electronic structures by rigidly shifting the top layer from the AA stacking along the high-symmetry [100] and $[1\bar{1}0]$ directions. The corresponding crystal structures at several representative stacking configurations during sliding are illustrated in Fig.~\ref{fig:FIG_S3}. The Fig~\ref{fig:FIG_S4} shows the evolution of stacking energy in antiferromagnetic states, defined as the energy difference between shifted configurations and AA stacking. The results indicate that the stackings with $n$/4 ($n$/6) fractional shifts exhibit local extrema along the [100] ($[1\bar{1}0]$) direction. In addition to the stacking energy, the spin and valley polarizations can also be effectively modulated by interlayer sliding.

To clearly illustrate the change of spin-splitting, we define two quantities, i.e., $\bigtriangleup E^{c(v)}_{spin}$ = $E^{c(v)}_{\uparrow}-E^{c(v)}_{\downarrow}$, which describe the nonrelativistic spin-splitting between the first and the second conduction (valence) bands near the Fermi level [see Fig.~\ref{fig:splitting}(e)]. Fig.~\ref{fig:splitting}(a) shows the $\bigtriangleup E^{c(v)}_{spin}$ for two opposite ferroelectric states ($\upuparrows$, $\downdownarrows$) as a function of fractional shifts along [100] and $[1\bar{1}0]$ directions. A prominent feature is that the sign of $\bigtriangleup E^{c(v)}_{spin}$ is opposite for two different ferroelectric states due to the contrasting spin-polarized band structures. With interlayer sliding, the sign of $\bigtriangleup E^{c}_{spin}$ can be switched for both ferroelectric states, whereas the sign for $\bigtriangleup E^{v}_{spin}$ remains unchanged. This sign change is closely related to the layer index. For example, Fig~\ref{fig:splitting}(e-h) illustrate the layer- and spin-resolved band structures for $\downdownarrows$ state with fractional shifts of 0, $\frac{1}{6}$, $\frac{1}{3}$ and $\frac{1}{2}$ along the [100] direction. Clearly, the sign of $\bigtriangleup E^{c}_{spin}$ changes between $\frac{1}{3}$ and $\frac{1}{2}$ when two spin states from different layers are exchanged. 

It is important to clarify the distinct mechanism of polarization switching in Nb$_3$I$_8$ compared to sliding ferroelectrics (such as V$X_2$ and VSi$_2X_4$), along with the corresponding differences in how spin polarization evolves under interlayer sliding. Both theoretical predictions and experimental observations have recently confirmed that Nb$_3$I$_8$ can undergo reversible switching between antiferroelectric and ferroelectric phases under the application of a vertical electric field~\cite{L-Feng2023,J-Hong2023}. The direction of the applied field determines the orientation of the resulting ferroelectric polarization, enabling control over two oppositely polarized ferroelectric states. Unlike sliding ferroelectrics, this ferroelectric reversal proceeds via structural reconfiguration of the top and bottom layers, which sequentially align into a uniformly polarized state, rather than through interlayer sliding. In sliding ferroelectric systems, a complete reversal of spin orientation across the entire structure can be achieved through interlayer sliding, as shown in V$X_2$ and VSi$_2X_4$ (see Fig.~\ref{fig:FIG_S2}). However, this mechanism does not apply to bilayer Nb$_3$I$_8$, as its polarization switching proceeds through internal structural changes rather than relative layer displacement. 

Moreover, the magnitudes of $\bigtriangleup E^{c(v)}_{spin}$ are also significantly affected by interlayer sliding. An interesting observation is that the curve shape of $\bigtriangleup E^{c(v)}_{spin}$ closely follows the evolution of the out-of-plane ferroelectric polarization component $P_z$, for both sliding along the high-symmetry [100] and $[1\bar{1}0]$ directions [seeing Fig. \ref{fig:splitting}(c)], indicating a positive correlation between spin-polarization and ferroelectric polarization. Additionally, the calculated maximum spin-splitting reaches up to 123 meV and 91 meV for valence and conduction bands, respectively. This FMs-like spin-splitting can be quantified as an analogue to the Zeeman effect. The induced effective magnetic field may be approximated as~\cite{RW-Zhang2024}
\begin{equation}
B_{c(v)}^{\textrm{eff}} = \dfrac{\bigtriangleup E^{c(v)}_{spin}}{g_{s}\mu_{B}\sigma_{z}}   \label{Zeeman}
\end{equation}
where $\sigma_{z} = 1$, $g_{s}$ is assumed to be 2, and $\mu_{B}$ is Bohr magneton. Applying Eq.~\ref{Zeeman}, the induced magnetic fields are estimated to be 1062 T and 786 T for the valence and conduction bands, respectively. These exceptionally high magnetic fields are of significant interest for the development of antiferromagnetic spintronics.

\begin{figure*}
	\includegraphics[width=2\columnwidth]{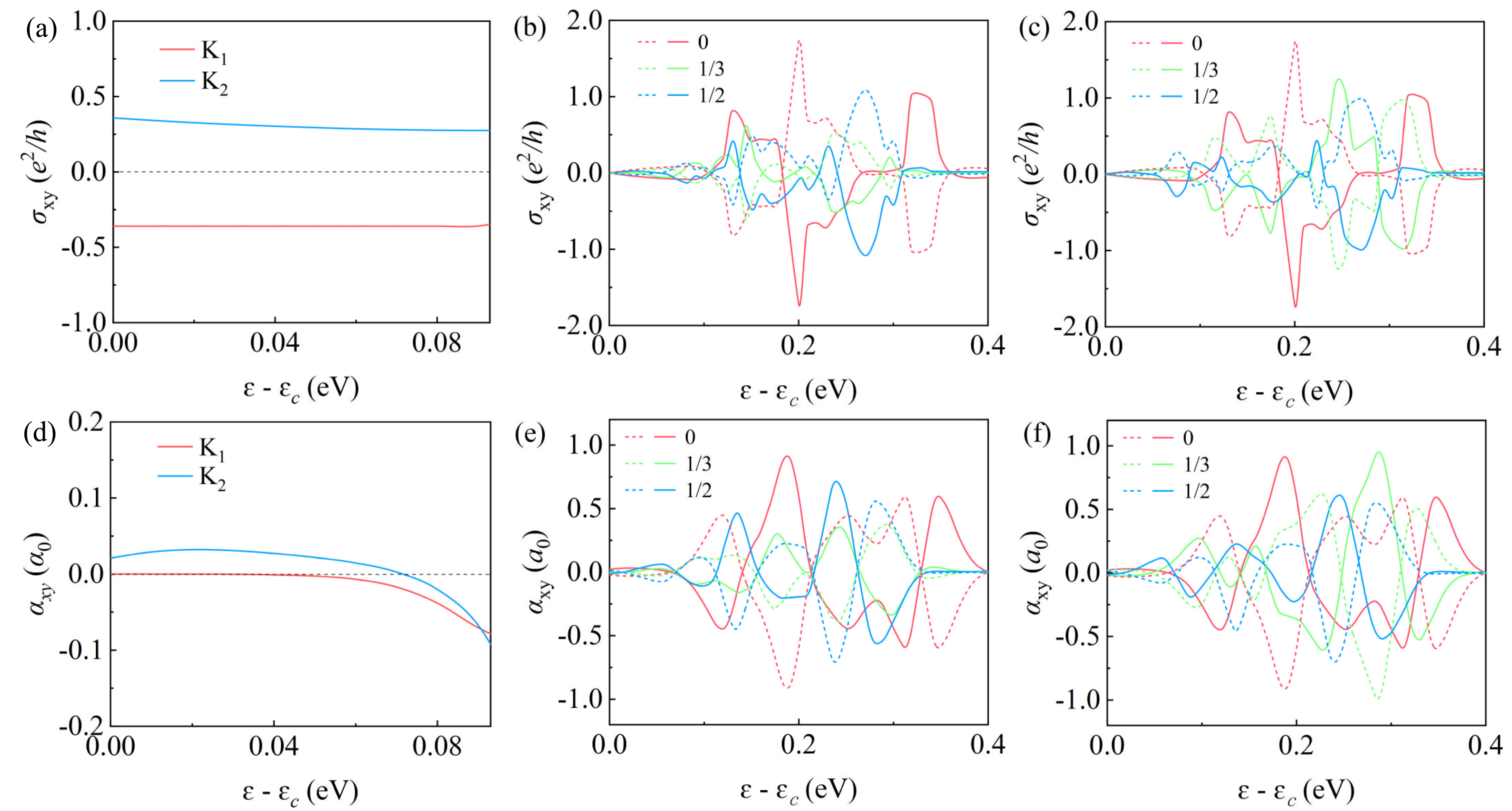}
	\caption{(Color online) Sign-Reversible and Size-Tunable VHE and VNE. (a) and (d) Valley-resolved $\sigma_{xy}$ and $\alpha_{xy}$ ($T$ = 100 K). (b-c) VHE as a function of Fermi energy for different fractional shifts along [100] and $[1\bar{1}0]$ directions, respectively. (e-f) Similar to (b-c) but with VNE at $T$ = 100 K. In panels (b-c) and (e-f), the solid and dashed lines represent two opposite ferroelectric states. The Fermi level $\varepsilon$-$\varepsilon_{c}$ are aligned with the conduction band minimum. The unit of VNE is defined as $\alpha_0 = k_B e/h$, where $k_B, e, h$ represent the Boltzmann constant,elementary charge, and Planck constant, respectively. }
	\label{fig:ane}
\end{figure*}

Considering the SOC effect, the valley-polarization is further induced when the N{\'e}el vector is oriented along the out-of-plane direction. Fig. \ref{fig:splitting}(i) presents the relativistic band structure of AA stacking, where the maximum valley-splitting ($\bigtriangleup E^{c,v}_{valley}$) is 24 meV and 93 meV for valence and conduction bands, respectively. Fig.~\ref{fig:splitting}(b) shows the $\bigtriangleup E^{c,v}_{valley}$ as a function of fractional shifts along [100] and $[1\bar{1}0]$ directions with the corresponding band structures shown in Fig.~\ref{fig:splitting}(i)-\ref{fig:splitting}(l). Both the conduction and valence band valley splittings exhibit clear variations as a function of interlayer sliding, reflecting the strong coupling between electronic structure and layer stacking. However, their responses differ in both magnitude and sign behavior. While the valence band valley splitting evolves monotonically with the out-of-plane ferroelectric polarization, the conduction band valley splitting shows a sign reversal that is not mirrored in the polarization trend. This contrasting behavior originates from their distinct layer-resolved orbital characters. The conduction band valleys are primarily composed of electronic states localized on different layers, making them highly sensitive to relative interlayer displacement. In contrast, the valence band valleys arise mainly from the same layer, and thus respond more directly to changes in the overall electric polarization. As a result, during sliding, the conduction band valleys undergo a reversal in energy order due to the interchange of dominant layer contributions (see Fig.~\ref{fig:FIG_S5}), whereas the valence band valley splitting follows a smooth, unidirectional trend consistent with polarization evolution.

 Furthermore, we also find that although the total magnetization remains zero and the sublattice magnetizations are nearly unchanged during interlayer sliding, the spin polarization projected onto the electronic states at the valence band maximum and conduction band minimum evolves noticeably and exhibits a clear and monotonic correlation with the variation of the out-of-plane ferroelectric polarization (see Fig.~\ref{fig:splitting}(d)). Such ferroelectrically controllable spin-valley behavior is reminiscent of earlier findings in nonmagnetic ferroelectric transition-metal oxide heterostructures~\cite{Yamauchi2015}. These findings further establish electric polarization as an active tuning parameter for spin and valley degrees of freedom in FE-AFMs. Slidetronics thus enables dynamic and continuous control of spin, valley, and ferroelectric polarization simultaneously. This approach is not limited to Nb$_3$I$_8$ bilayers but can be generalized to other FE-AFMs with an out-of-plane electric dipole moment.

The nonzero spin-valley polarization in FE-AFMs opens up exciting possibilities for antiferromagnetic spintronic and valleytronic applications, such as VHE and VNE. The observable VHE and VNE refer to the generation of a net Hall current when electrons from distinct valleys experience opposite but unbalanced fictitious magnetic fields [i.e., Berry curvature $\Omega(\bold{k})$] under an electric field and temperature gradient. Fig.~\ref{fig:splitting}(m) shows the momentum-resolved $\Omega(\bold{k})$ for AA stacking. The hot spots of $\Omega(\bold{k})$ are primarily concentrated around two K valleys, with opposite signs and different magnitudes. With interlayer sliding, the $\Omega(\bold{k})$ always exhibits an antisymmetric distribution [see Fig.~\ref{fig:splitting}(n)-\ref{fig:splitting}(p)] due to the reduced magnetic symmetry compared to AA stacking. Moreover, the magnitudes of $\Omega(\bold{k})$ at two valleys can be dynamically tuned through sliding.

The switchability of Berry curvature is expected to significantly influence the VHE, VNE, and other spin-valley-related transport phenomena. The valley-resolved anomalous Hall conductivity ($\sigma_{xy}$) is determined by integrating the Berry curvature over a small region centered around each valley, while the valley-resolved anomalous Nernst conductivity ($\alpha_{xy}$) involves integrating the Berry curvature with a weighting factor near each valley [see Eqs.~\ref{eq:sigma}-\ref{eq:FnK}]. Fig.~\ref{fig:ane}(a) and ~\ref{fig:ane}(d) present the variation of valley-resolved $\sigma_{xy}$ and $\alpha_{xy}$ for the AA-stacked Nb$_3$I$_8$ bilayer at the bottom of conduction band. The case for valance band edge is presented in Fig.~\ref{fig:FIG_S6}. Apparently, the $\sigma_{xy}$ at the two K valleys exhibit opposite response but with different magnitudes, in line with the behavior of $\Omega(\bold{k})$. In contrast, for $\alpha_{xy}$, the contribution from the K$_1$ valley is nearly zero, while the K$_2$ valley contributes nonzero values. This difference arises because the $\sigma_{xy}$ and $\alpha_{xy}$ are intimately related to
each other through the generalized Mott formula\cite{D-Xiao2006},
\begin{equation}\label{eq:ane}
\alpha _{xy} =-\frac{1}{e} \int d\varepsilon \frac{\partial f}{\partial \mu } \sigma _{xy} \frac{\varepsilon -\mu }{T} ,
\end{equation}
where $e$, $\varepsilon$, $f$, $\mu$ and $T$ are the elementary charge, energy, Fermi-Dirac distribution function, chemical potential, and temperature, respectively. In the low-temperature limit, this expression simplifies to the Mott relation, 
linking the $\alpha_{xy}$ to the energy derivative of the $\sigma_{xy}$,
\begin{equation}\label{eq:mott}
\alpha _{xy} =-\frac{\pi ^{2}k^{2} _{B} T }{3} {\sigma _{xy} }' \left ( \varepsilon  \right ) ,
\end{equation}
From Fig.~\ref{fig:ane}(a), it is evident that the $\sigma_{xy}$ for K$_1$ (K$_2$) valley presents a plateau (negative slope) with the change of energy. Consequently, the $\alpha_{xy}$ for K$_1$ (K$_2$) valley, shown in Fig.~\ref{fig:ane}(d), is vanishing (positive). At the high-energy region, the situation may become more complicated, as the contributions of Berry curvature may arise not only from the two valleys but also from other momentum regions.

The uncompensated $\sigma_{xy}$ and $\alpha_{xy}$ at two K valleys, which are driven by electric field and temperature gradient, will force a net transverse Hall current, respectively. Fig.~\ref{fig:ane}(b-c) and \ref{fig:ane}(e-f) show the slidetronic tunable VHE and VNE  along the [100] and $[1\bar{1}0]$ directions, respectively. It is clear that these quantities can be dynamically switched via interlayer sliding. Furthermore, the VHC and VNC can be sign-reversible by switching the ferroelectric states as their odd properties with respect to $\mathcal{TP}$ symmetry. In fact, the influence of ferroelectric polarization have been previously witnessed in phenomena such as anomalous Hall effect, magneto-optical effect, and nonlinear transports~\cite{FR-Fan2017,Yananose2022,N-Ding2023,YY-Feng2023,L-Feng2023,JQ-Feng2024,WC-Sun2024,RC-Xiao2022}. But the flexible control of spin-valley polarization and the resulting transport properties remain relatively unexplored. Moreover, one can also find that the induced $\sigma_{xy}$ and $\alpha_{xy}$ exhibit remarkably high values, e.g., $\sigma_{xy}$ = 1.8 $e^2/h$ (about 240.8 S/cm) and $\alpha_{xy}$ = 0.9 $a_{0}$ (about 2.1 A/Km) at 100 K with appropriate electron doping. These magnitudes are camparable and even larger than those in the well-known 2D ferromagnetic Fe$_n$GeTe$_2$ bilayer ($n$ = 3, 4, 5)\cite{XX-Yang2021b}, 1T-CrTe$_{2}$~\cite{XX-Yang2021}, Cr$XY$ ($X$ = S, Se, Te; $Y$ = Cl, Br, I)~\cite{XX-Yang2022}, and 3D bcc Fe~\cite{XK-Li2017}. The large valley transport arises from the giant effective magnetic field and strong SOC, sharing a similar physical mechanism with the ferromagnetic materials.

\section{Summary}

In conclusion, we introduce a novel class of $\mathcal{TP}$-broken layered ferroelectric antiferromagnets (FE-AFMs) as a platform for spontaneous spin-valley polarization, combining the favorable features of spin-polarized altermagnets and valley-polarized $\mathcal{TP}$-symmetric AFMs. FE-AFMs are widely present in various layered antiferromagnets, such as Nb$_3X_8$ ($X$ = Cl, Br, I), V$X_2$ ($X$ = S, Se), and VSi$_2X_4$ ($X$ = N, P). These materials exhibit unique characteristics, including global layer-dependent spin-polarization and uncompensated density of states, which arise from their intrinsic ferroelectric polarization. The ability to tune both the sign and magnitude of spin-valley polarization in FE-AFMs opens up exciting prospects for sign-reversible and size-controllable valley Hall and Nernst effects, along with other valley-dependent transport phenomena. Our findings underscore the potential of FE-AFMs for advancing spintronic and valleytronic technologies, paving the way for the development of multifunctional devices that capitalize on the combined benefits of both spintronic and valleytronic functionalities.

\begin{acknowledgments}
This work is supported by the National Natural Science Foundation of China (Grants Nos. 12304066, 12404052, 12474012, and 12074154), the Basic Research Program of Jiangsu (Grant No. BK20230684 and BK20241049), the Natural Science Fund for Colleges and Universities in Jiangsu Province (Grant No. 23KJB140008 and 24KJB140011), Six Talent Peaks Project and 333 High-level Talents Project of Jiangsu Province.
 \end{acknowledgments}
 
 	\appendix
	\section{The details of first-principles calculations}\label{appendix1}

The first-principles calculations are performed using the  Vienna \textit{ab initio} simulation package (\textsc{vasp})~\cite{Kresse1996a,Kresse1996} with the PAW method~\cite{Bloechl1994} and GGA-PBE functional~\cite{Perdew1996}. To account for the correlation effects in the $d$-orbitals of Nb atoms (for Nb$_3$I$_8$), V atoms (for VS$_2$ and VSi$_2$N$_4$), and Mn atoms (for MnPSe$_3$), the GGA + \textit{U} method by Dudarev $et \ al$.~\cite{Dudarev1998} was applied with an effective Hubbard parameter of \textit{U} = 3.0~\cite{JQ-Feng2024}, 3.0~\cite{XG-Liu2020,XD-Zhou2021b}  and 5.0 eV~\cite{X-Li2013}, respectively. For V$_2$Se$_2$O, the GGA + \textit{U} method by Liechtenstein $et \ al$~\cite{Liechtenstein1995} was used with Hubbard $U$ of 5.1 eV and Hund$^\prime$s rule coupling $J$ of 0.8 eV~\cite{HY-Ma2021}. A plane-wave cutoff energy of 500 eV was used, and the $k$-point mesh was set to 11$\times$11$\times$1, 35$\times$35$\times$1, 21$\times$21$\times$1, 13$\times$13$\times$1 and  21$\times$21$\times$1 for Nb$_3$I$_8$, VS$_2$, VSi$_2$N$_4$, MnPSe$_3$, and V$_2$Se$_2$O, respectively. The energy and force convergence criterion were set to be $10^{-6}$ eV and $10^{-2}$ eV/\AA, respectively. A vacuum layer of at least 15 {\AA} is used to prevent interactions between periodic images. The interlayer Van der Waals interactions were incorporated using the DFT-D3 method~\cite{Grimme2010,Grimme2011}. The spin textures are drawn using the VASPKIT software~\cite{V-Wang2021}.

The maximally localized Wannier functions for the $p$-orbitals of I atoms and $d$-orbitals of Nb atoms were constructed using the \textsc{WANNIER90} package~\cite{Mostofi2008}, with an 8$\times$8$\times$1 $k$-point mesh. The transverse electronic and thermoelectric transport properties were subsequently calculated using a tight-binding Hamiltonian on the basis of Wannier functions. The intrinsic $\sigma_{xy}$ and $\alpha_{xy}$ were determined using Berry phase theory and the Kubo formula with a $k$-mesh of 501$\times$501$\times$1~\cite{D-Xiao2010,Kubo1957,WX-Feng2016},
\begin{eqnarray}\label{eq:sigma}
\sigma_{xy}=-\frac{e^{2}}{\hbar} \sum_{n} \int \frac{d^{2} k}{(2 \pi)^{2}} \Omega_{xy}^{n}(\boldsymbol{k}) f_{n}(\boldsymbol{k}), 
\end{eqnarray}
\begin{eqnarray}\label{eq:alpha}
\alpha_{xy}=-\frac{e^{2}}{\hbar} \sum_{n} \int \frac{d^{2} k}{(2 \pi)^{2}} \Omega_{xy}^{n}(\boldsymbol{k}) F_{n}(\boldsymbol{k}), 
\end{eqnarray}
where $\Omega_{xy}^{n}(\boldsymbol{k})$ is band-resolved Berry curvature, expressed as:
\begin{equation}\label{eq:berry}
 \Omega_{xy}^{n}(\boldsymbol{k})=-\sum_{n^{\prime} \neq n} \frac{2 \operatorname{Im}\left[\left\langle\psi_{n \boldsymbol{k}}\left|\hat{v}_{x}\right| \psi_{n^{\prime} \boldsymbol{k}}\right\rangle\left\langle\psi_{n^{\prime} \boldsymbol{k}}\left|\hat{v}_{y}\right| \psi_{n \boldsymbol{k}}\right\rangle\right]}{\left(\omega_{n^{\prime} \boldsymbol{k}}-\omega_{n \boldsymbol{k}}\right)^{2}} 
\end{equation}
with $\hat{v}_{x,y}$, $\psi_{n\boldsymbol{k}}$, $\varepsilon_{n\boldsymbol{k}}$, and $f_{n}$($\boldsymbol{k}$) are velocity operators, eigenvectors, eigenvalues, and Fermi-Dirac distribution function, respectively. The $F_{n}(k)$ is defined as
\begin{eqnarray}\label{eq:FnK}
\begin{aligned}
F_{n}(\boldsymbol{k})= & -\frac{1}{e T}\left[\left(\varepsilon_{n \boldsymbol{k}}-\mu\right) f_{n}(\boldsymbol{k})\right. \\
& \left.+k_{B} T \ln \left(1+e^{-\left(\varepsilon_{n \boldsymbol{k}}-\mu\right) / k_{B} T}\right)\right] 
\end{aligned}
\end{eqnarray}
where $T$ is the temperature, $\mu$ the chemical potential, and $k_B$ the Boltzmann constant. 

\makeatletter
\@addtoreset{figure}{section}  
\makeatother

\section{Supplemental Figure}
\label{appendix2}

\setcounter{figure}{0} 
\renewcommand{\thefigure}{S\arabic{figure}}

\begin{figure}[th]
	\includegraphics[width=\columnwidth]{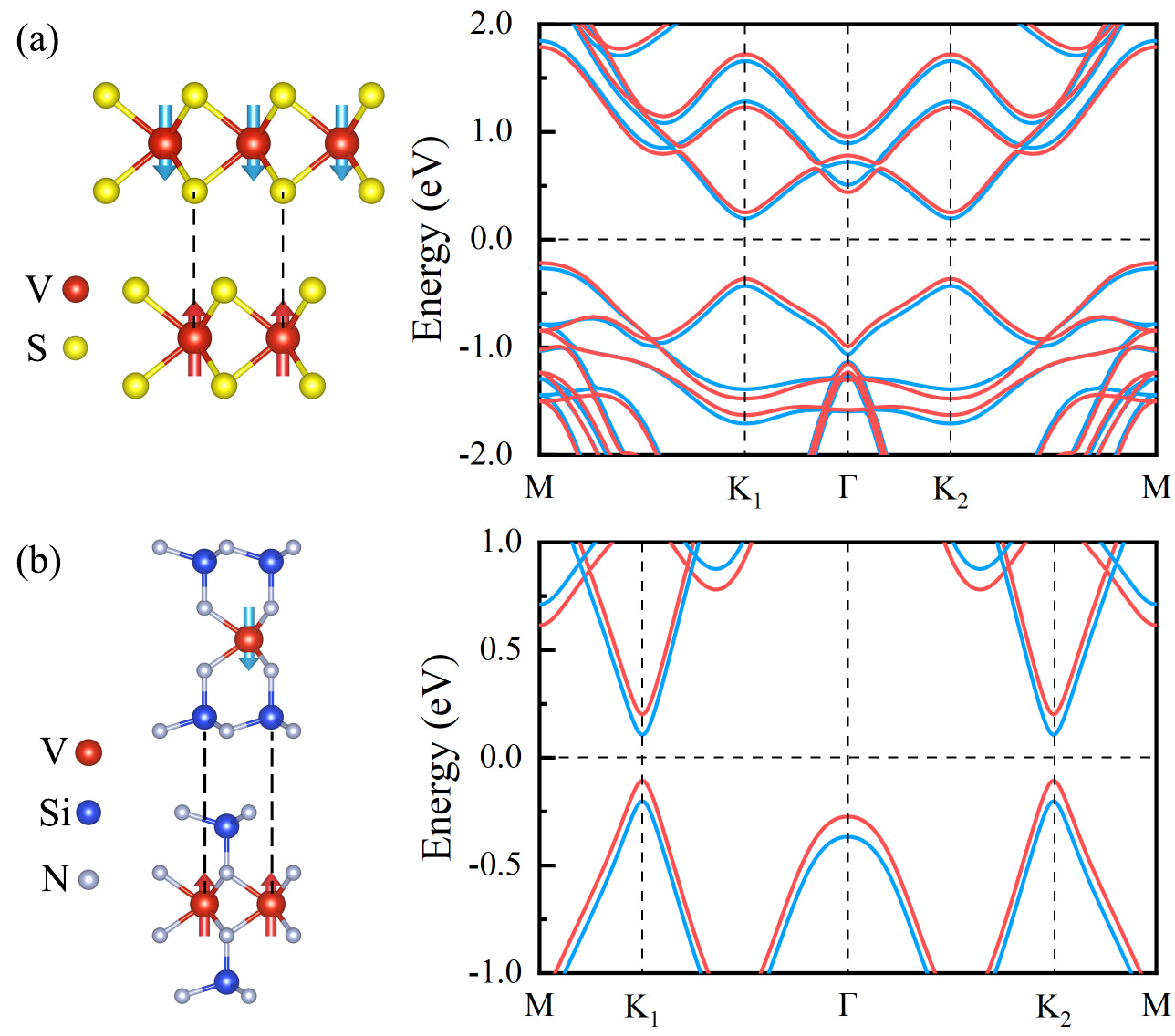}
	\caption{(a)-(b) Magnetic structure and spin-polarized band structures for bilayer VS$_2$ and VSi$_2$N$_4$.}
	\label{fig:FIG_S1}
\end{figure}

\begin{figure}[th]
	\includegraphics[width=\columnwidth]{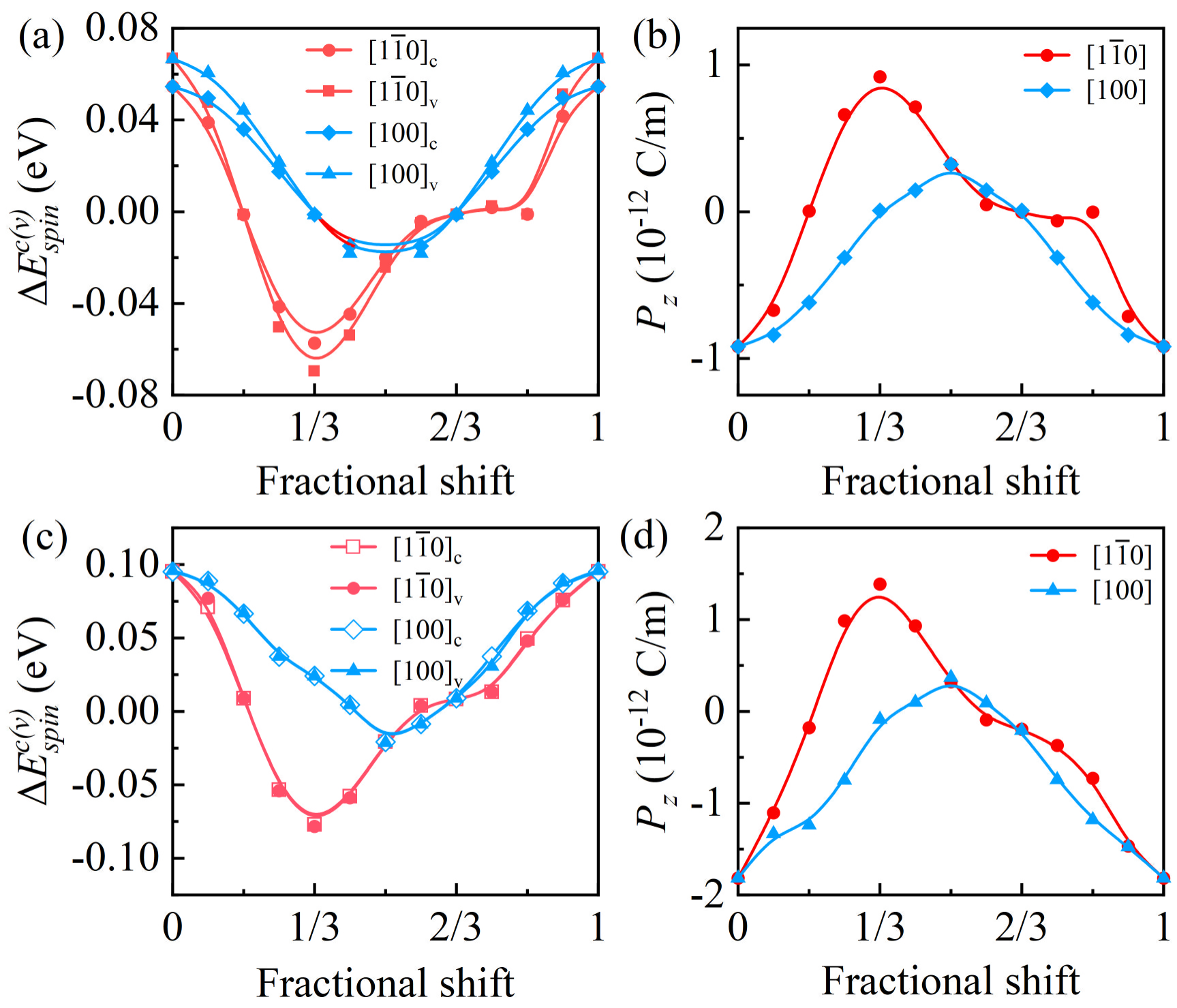}
	\caption{Spin-splitting and out-of-plane ferroelectric polarization component as a function of lateral shift along the [100] and [1$\bar{1}$0] directions for VS$_2$ (a-b) and VSi$_2$N$_4$ (c-d). We note that VSe$_2$ and VSi$_2$P$_4$, which share the same crystal structure as VS$_2$ and VSi$_2$N$_4$, are expected to exhibit similar behavior. Interestingly, the variation in spin splitting closely mirrors the evolution of the out-of-plane polarization component ($P_z$), in agreement with our observations in Nb$_3$I$_8$. This parallel behavior highlights the general nature of spin-valley coupling in FE-AFM systems. However, a notable distinction arises between Nb$_3$I$_8$ and materials like VS$_2$ and VSi$_2$N$_4$. In the latter, the ferroelectric polarization flips its sign during interlayer sliding, characteristic of sliding ferroelectrics. In contrast, Nb$_3$I$_8$ retains a fixed polarization direction throughout the sliding process, owing to its inherent structural symmetry. This difference accounts for the synchronized reversal of spin splitting with polarization in VS$_2$ and VSi$_2$N$_4$.}
	\label{fig:FIG_S2}
\end{figure}

\begin{figure}[th]
	\includegraphics[width=\columnwidth]{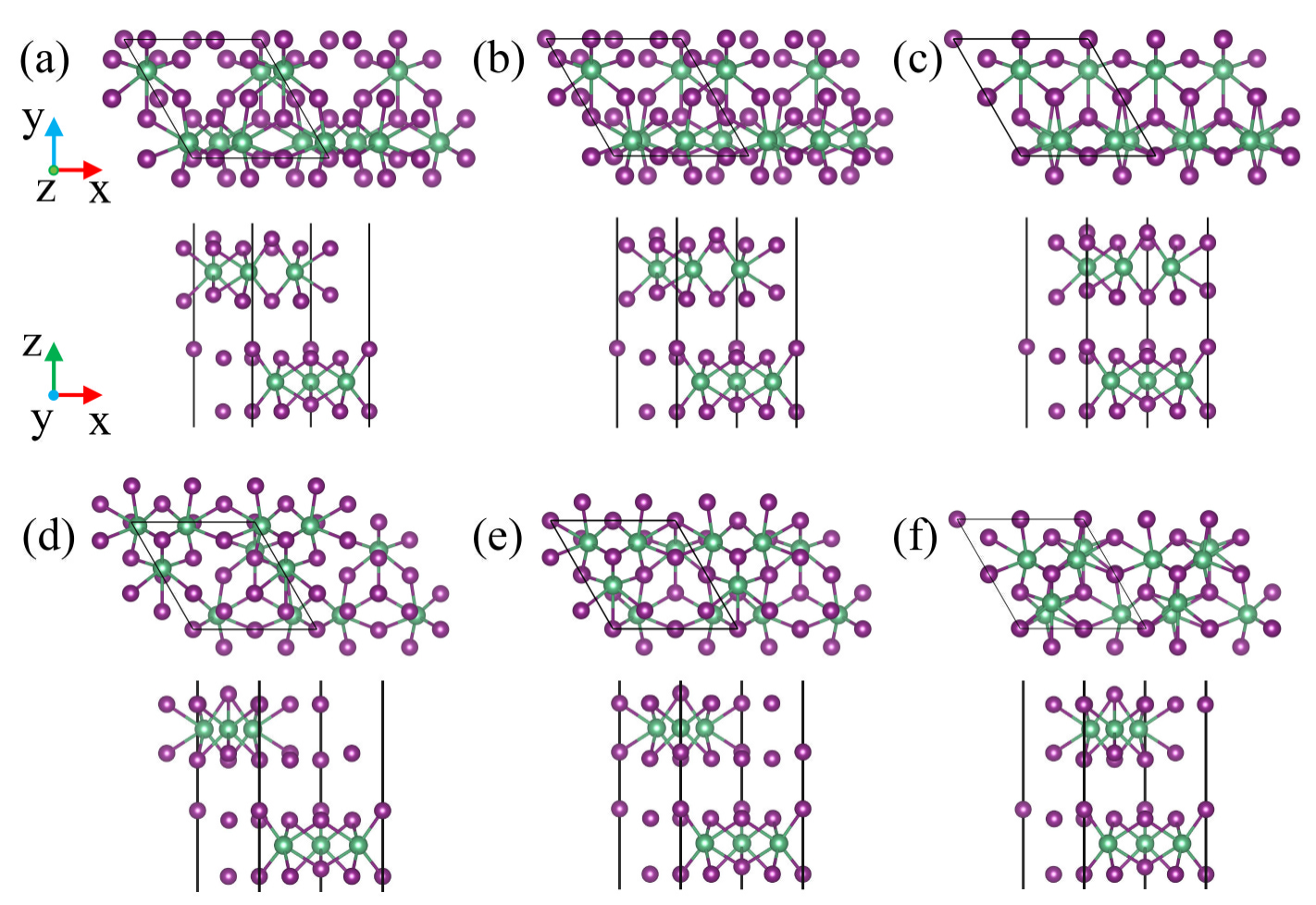}
	\caption{(a-c) and (d-f), top and side views of bilayer Nb$_3$I$_8$ with 1/6, 1/3 and 1/2 fractional shifts from AA stacking along the [100] and [1$\bar{1}$0] directions, respectively.}
	\label{fig:FIG_S3}
\end{figure}

\begin{figure}[th]
	\includegraphics[width=0.8\columnwidth]{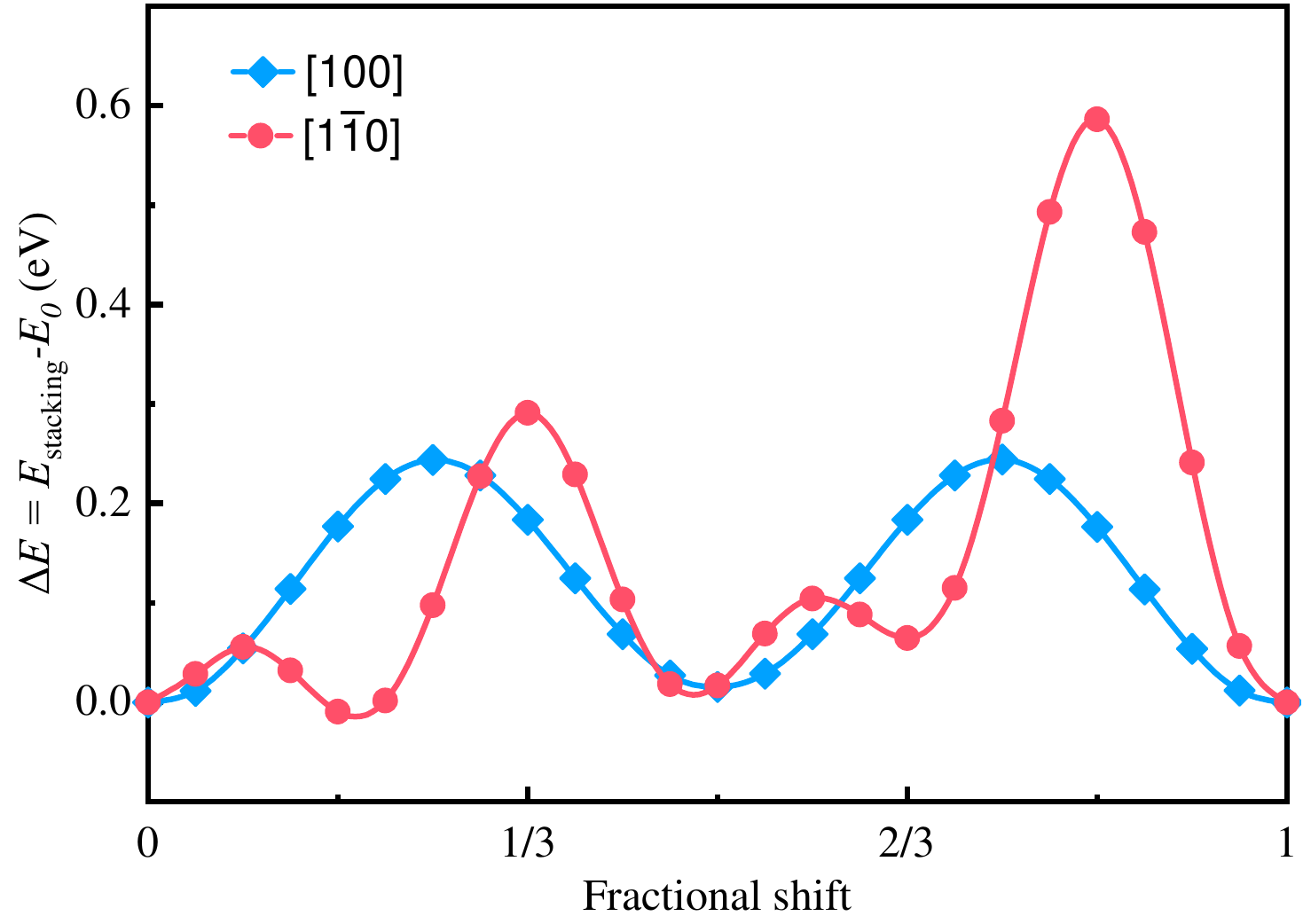}
	\caption{Stacking energy for bilayer Nb$_3$I$_8$ as a function of lateral shift along the [100] and [1$\bar{1}0$] direction.}
	\label{fig:FIG_S4}
\end{figure}

\begin{figure}[th]
	\includegraphics[width=\columnwidth]{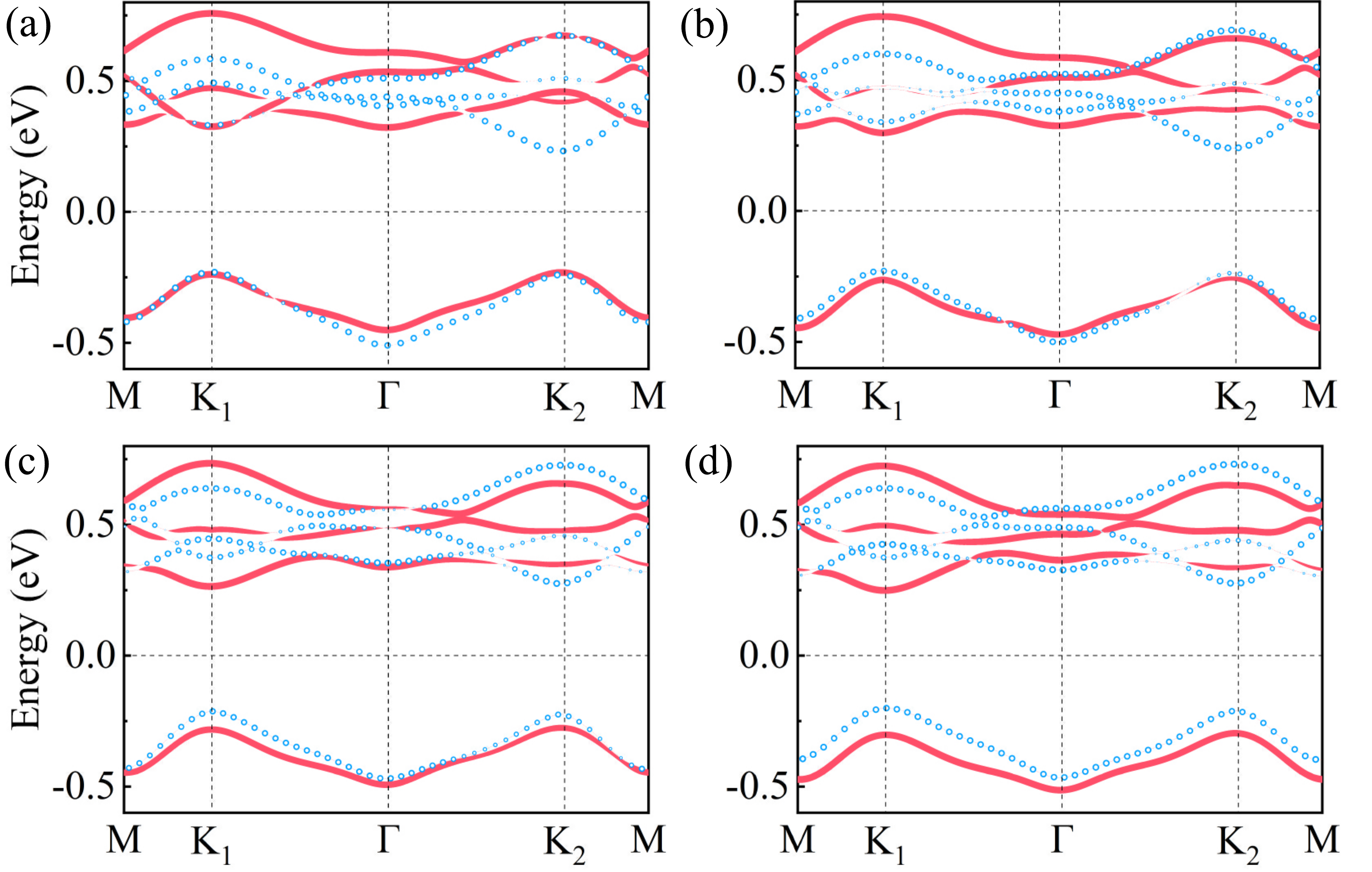}
	\caption{(a-d) Layer-resolved relativistic band structures for bilayer Nb$_3$I$_8$ with fractional interlayer shifts of 0, $\frac{1}{6}$, $\frac{1}{3}$ and $\frac{1}{2}$ along the [100] direction, respectively. The red solid lines denote the contribution from bottom layer, while blue dashed lines represent those from the top layer.}
	\label{fig:FIG_S5}
\end{figure}

\begin{figure}[th]
	\includegraphics[width=0.8\columnwidth]{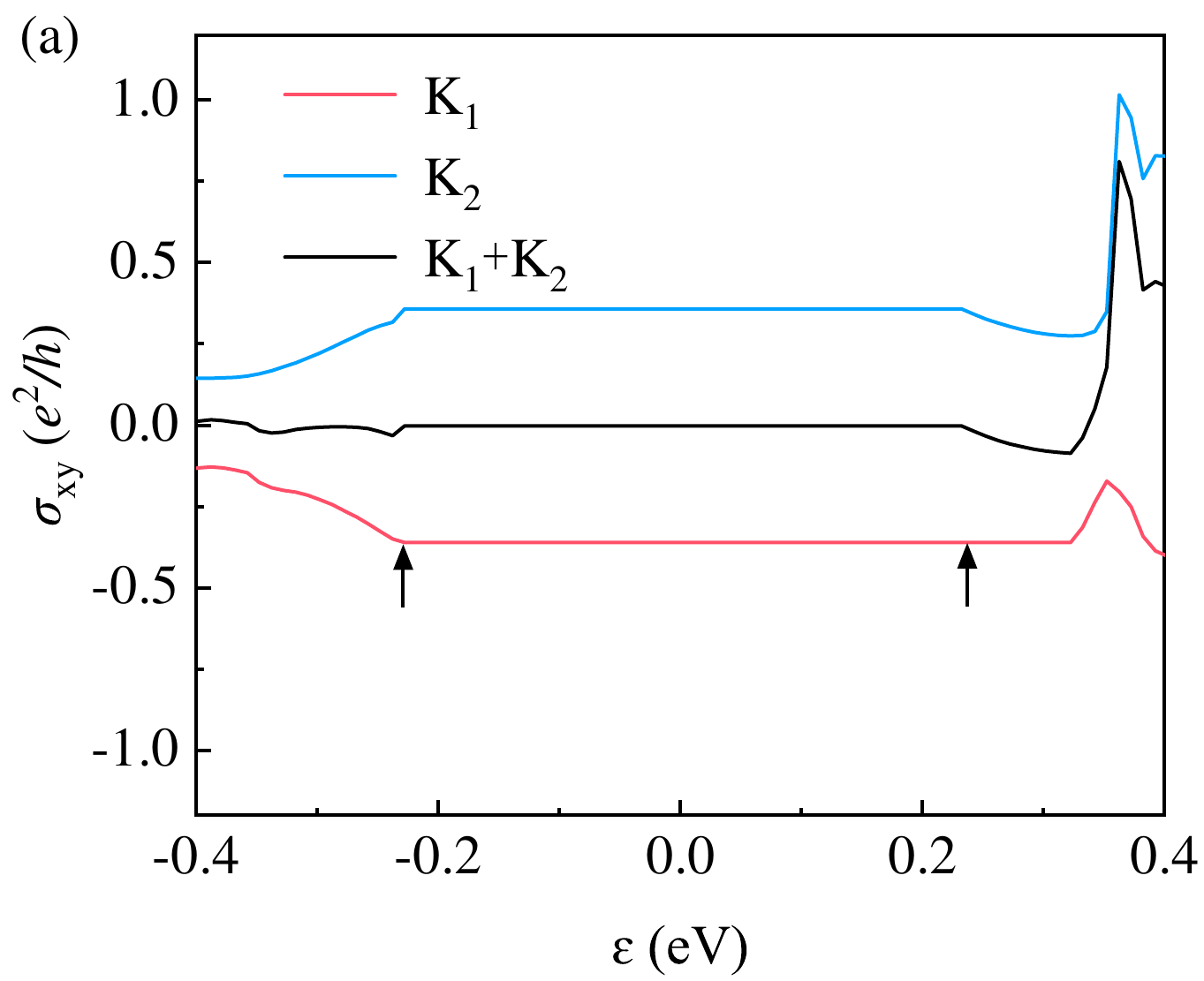}
	\caption{Valley-resolved (K$_1$, K$_2$) and total (K$_1$+K$_2$) anomalous Hall conductivity as a function of energy. The black arrows indicating the valence band maximum and the conduction band minimum. One can see that  the anomalous Hall conductivity exhibits relatively small values near the valence band edge, while significantly larger values are observed near the conduction band edge.}
	\label{fig:FIG_S6}
\end{figure}

\clearpage


%

\end{document}